\documentclass[12pt,preprint]{aastex}

\def\ion#1#2{#1\,{\sc\romannumeral #2}}

\usepackage{graphicx}

\shorttitle{Temperatures in the Solar Corona}
\shortauthors{Warren \& Brooks}

\begin{document}


\title{The Temperature and Density Structure of the Solar Corona. I. Observations of the
  Quiet Sun with the EUV Imaging Spectrometer (EIS) on \textit{Hinode}}

\author{Harry P. Warren and David  H. Brooks\altaffilmark{1}}

\affil{Space Science Division, Naval Research Laboratory, Washington,
  DC 20375}

\altaffiltext{1}{College of Science, George Mason University, 4400
University Drive, Fairfax, VA 22030}


\begin{abstract}
  Measurements of the temperature and density structure of the solar corona provide
  critical constraints on theories of coronal heating. Unfortunately, the complexity of
  the solar atmosphere, observational uncertainties, and the limitations of current atomic
  calculations, particularly those for Fe, all conspire to make this task very
  difficult. A critical assessment of plasma diagnostics in the corona is essential to
  making progress on the coronal heating problem.  In this paper we present an analysis of
  temperature and density measurements above the limb in the quiet corona using new
  observations from the EUV Imaging Spectrometer (EIS) on \textit{Hinode}.  By comparing
  the Si and Fe emission observed with EIS we are able to identify emission lines that
  yield consistent emission measure distributions. With these data we find that the
  distribution of temperatures in the quiet corona above the limb is strongly peaked near
  1\,MK, consistent with previous studies. We also find, however, that there is a tail in
  the emission measure distribution that extends to higher temperatures. EIS density
  measurements from several density sensitive line ratios are found to be generally
  consistent with each other and with previous measurements in the quiet corona. Our
  analysis, however, also indicates that a significant fraction of the weaker emission
  lines observed in the EIS wavelength ranges cannot be understood with current atomic
  data.
\end{abstract}



\section{Introduction}

The origin of the high temperature plasma that permeates the solar corona has defied
understanding for many decades. In principal, the thermal structure of the solar corona
holds many clues to the physical processes that convert magnetic energy into thermal
energy. For example, it has been proposed that the corona is heated by frequent bursts of
magnetic reconnection called nanoflares \citep[e.g.][]{parker1972,parker1983}. In this
model turbulent motions in the solar photosphere lead to the constant tangling and
braiding of the magnetic fields that rise up into the corona. The dissipation of this
topological complexity leads to the release of energy on very small spatial scales. This
suggests that the corona should be composed of many fine loops that are in various stages
of heating and cooling. Thus the distribution of temperatures and densities is a critical
constraint on the frequency, duration, and magnitude of the heating events that give rise
to the high temperature corona.

Unfortunately, determining the distribution of temperatures and densities in the corona is
a nontrivial problem. The solar corona is highly structured and highly dynamic, making it
difficult to isolate individual structures. Obtaining accurate atomic calculations is also
a problem. Because of its relatively high elemental abundance, emission from Fe has been
the focus of many recent solar instruments. Interpreting observations from imaging
instruments such as SXT/\textit{Yohkoh}, EIT/\textit{SOHO}, TRACE, EUVI/\textit{STEREO},
XRT/\textit{Hinode}, and the upcoming AIA/\textit{SDO} depends critically on accurate
atomic calculations for Fe. The accuracy of the available atomic calculations for this
complex atom, however, is often unclear.

The launch of the EUV Imaging Spectrometer (EIS) on \textit{Hinode} has greatly expanded
spectroscopic observations of the solar corona. EIS combines a broad temperature coverage
(\ion{Fe}{8}--\ion{Fe}{17}, \ion{Fe}{22}--\ion{Fe}{24}) with relatively high spatial
(1\arcsec) and spectral (22\,m\AA) resolution, allowing the temperature and density
structure of the corona to be examined in great detail. Of particular interest are the
properties of coronal loops observed in solar active regions and flares. Current models of
active region loops suggest that the observed densities and temperatures are signatures of
non-equilibrium processes, and EIS, with its advanced diagnostic capabilities, provides
much stricter observational constraints on physical models. Recent work on the properties
of active region loops, for example, has suggested that active region loops near 1\,MK
have narrow distributions of temperature, high densities, and relatively small filling
factors \citep{warren2008b}.

Since many of the plasma diagnostics provided by EIS are based on Fe emission lines it is
important to assess them critically and compare results with previous measurements. Some
initial results have been alarming. For example, EIS spectroheliograms in \ion{Fe}{8} and
\ion{Si}{7} are nearly identical, suggesting a very similar temperature of formation
\citep{young2007}. The respective peaks in the ionization fractions, however, are separated
by over 0.2\,MK. This calls into question how accurately temperatures can be measured
using the Fe emission observed in the EIS wavelength ranges.

In this paper we present an analysis of EIS observations above the quiet solar limb. Many
previous observations of the quiet corona have suggested that the distribution of
temperatures is very narrow, almost isothermal (e.g.,
\citealt{raymond1997,feldman1998,feldman1999a,warren1999c,allen2000,landi2002,warren2002a}). By
comparing Fe and Si emission observed with EIS we are able to identify plasma diagnostics
that are both self-consistent and in agreement with earlier results. Fortunately, most of
the useful diagnostics that are identified are the strongest emission lines that can be
observed with EIS. However, our analysis also shows that a significant fraction of the
observed emission lines cannot be understood with current atomic models.

In this work we also apply a Monte Carlo Markov Chain (MCMC) emission measure algorithm
\citep{kashyap1998,kashyap2000} to the observed EIS spectra and find that the distribution
of temperatures in the quiet corona is more complicated than previously thought. The
emission measure is sharply peaked near 1\,MK, as was found in earlier studies. We also
find, however, a tail in the emission measure distribution that extends to higher
temperatures. A high temperature component to the emission measure distribution is a
critical element of impulsive coronal heating models
\citep[e.g.,][]{cargill2004,patsourakos2008}.

Previous measurements of electron densities in the quiet corona above the limb have
yielded values of $\log n_e\sim8.3$ \citep[e.g.,][]{doschek1997b}. EIS has several density
sensitive line ratios that are useful in the quiet corona. We find that these ratios are
all generally consistent with each other and with previous results, although there is
considerable dispersion in the densities inferred from EIS.

Finally, we also discuss the potential for measuring relative abundances in the corona
using the \ion{S}{10} 264.233\,\AA\ and \ion{O}{6} 184.117\,\AA\ lines observed with EIS.

\section{Instrumentation and Data Reduction}

The EIS instrument on \textit{Hinode} produces stigmatic spectra in two wavelength ranges
(171--212\,\AA\ and 245--291\,\AA) with a spectral resolution of 0.0223\,\AA\
\citep{culhane2007,korendyke2006}. There are 1\arcsec\ and 2\arcsec\ slits as well as
40\arcsec\ and 266\arcsec\ slots available. The slit-slot mechanism is 1024\arcsec\ long
but a maximum of 512 pixels on the CCD can be read out at one time. Solar images can be
made using one of the slots or by stepping one of the slits over a region of the
Sun. Telemetry constraints generally limit the spatial and spectral coverage of an
observation. For this work we focus on the results from a special observing sequence
(\verb+HPW_FULLCCD_001+) that returned the entire wavelength range of the CCD over a small
region on the Sun ($128\arcsec\times128\arcsec$).  At each position in the raster a 90\,s
exposure is taken.

\begin{figure*}
\centerline{%
\includegraphics[clip,scale=1.0]{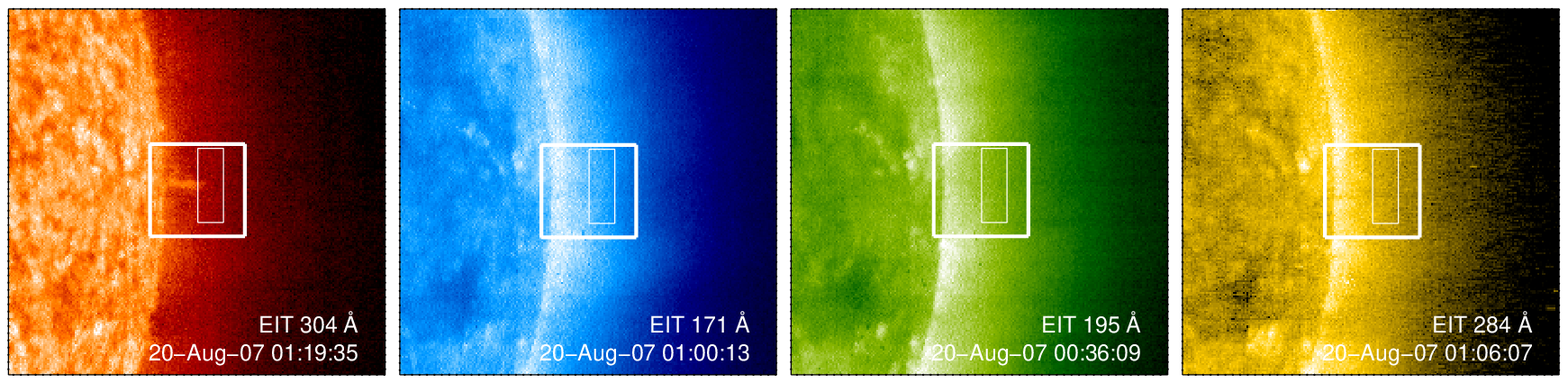}}
\centerline{%
\includegraphics[clip,scale=1.0]{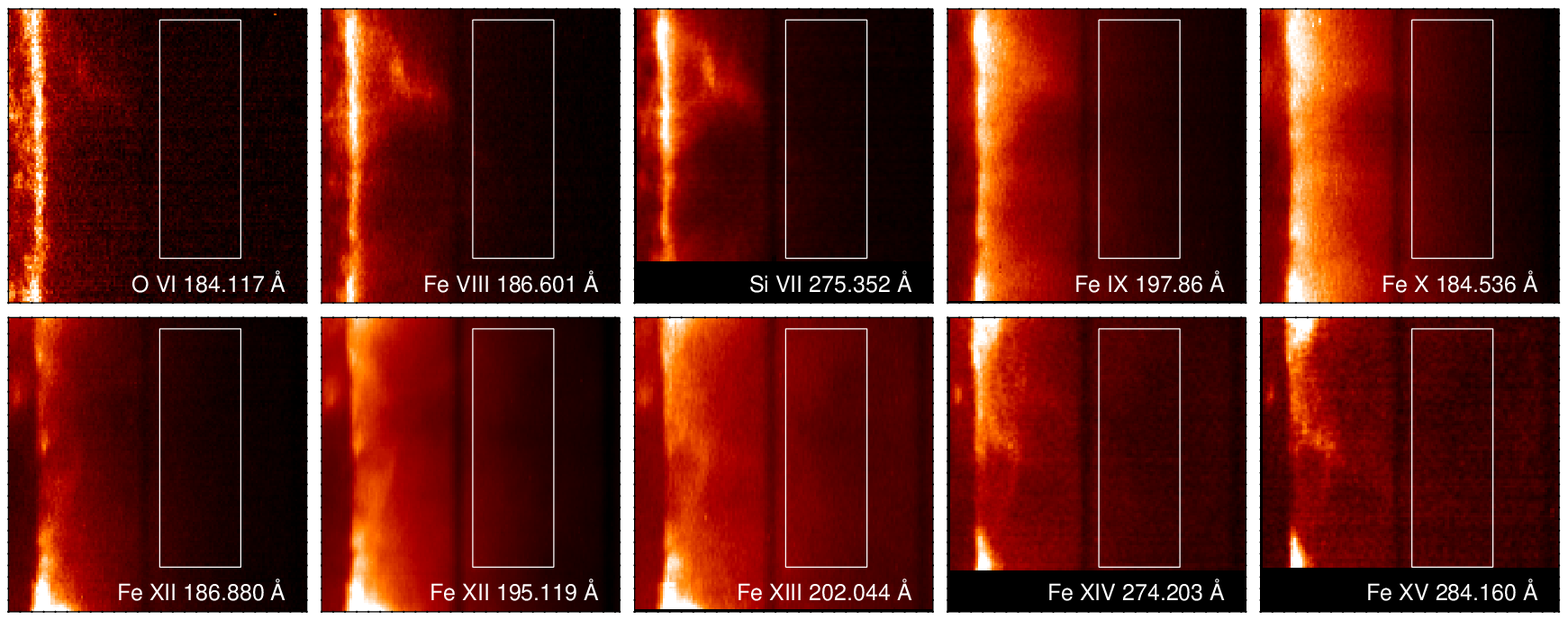}}
\caption{(\textit{top panels}) EIT images taken during the EIS full CCD observations. The
  boxes indicate the EIS field of view and the region used for computing the average
  spectra. (\textit{bottom panels}) EIS spectroheliograms in selected emission lines. The
  dark vertical band in each image is due to atmospheric absorption during a brief orbital
  eclipse.}
\label{fig:eit}
\end{figure*}

The data for this study were taken between August 19, 2007 23:47 UT and August 20, 2007
03:01 at the west limb. Context images from the EUV Imaging Telescope on \textit{SoHO}
\citep{delaboudiniere1995} are shown in Figure~\ref{fig:eit} and indicate that the
observed region was quiet during this time. Inspection of EIT 195\,\AA\ movies during this
period show no indication of significant coronal activity. The EIT 304\,\AA\ image does
indicate the presence of some cool material at low heights in the corona.

These EIS data were processed using standard algorithms to remove the CCD pedestal, dark
current, ``cosmic ray'' spikes, and warm pixels. The data numbers recorded in each pixel
were also converted to physical units (erg cm$^{-2}$ s$^{-1}$ sr$^{-1}$ \AA$^{-1}$). For each
intensity value an uncertainty is also calculated. This uncertainty includes counting
statistics and read noise, but not the uncertainty associated with the absolute
calibration. The implications of this will be discussed in later sections.

There are several instrumental effects that impact our analysis. The first is the
oscillation in the line centroids. This oscillation, which is about 2 spectral pixels in
magnitude ($\sim0.04$\,\AA) over a period of about 90 minutes, is believed to be due to
changing thermal conditions on the spacecraft during an orbit. To correct for this we
assume that the average Doppler shift in the \ion{Fe}{12} 195.119\,\AA\ line averaged
along the slit is zero at each slit position. Another important effect is the spatial
offset between the two detectors. Because of a misalignment between the two CCDs there is
a vertical offset of approximately 18 pixels between images taken in the different
channels. In data taken before 2008 August 24 there is also an offset of 1--2 pixels in
the solar-X direction.

After these instrumental effects are accounted for we can determine the line intensity at
each spatial position by calculating either moment of the line profile or a Gaussian
fit. For making context rasters we use moments and EIS rasters in selected lines are shown
in Figure~\ref{fig:eit}. These rasters also indicate the presence of some cool material at
low heights in the corona.

Since our objective is to analyze observations from the quiet corona with very high signal
to noise we have computed a spectrum averaged over a $35\arcsec\times104\arcsec$ region
that lies above this cool material. The center of this region is about 70\arcsec\ above
the limb or about 1.07$R_\odot$. In constructing this average spectrum we have ignored any
pixel that has been marked as a warm pixel or as a cosmic ray impact. The intensity in
each spectral pixel is computed from the average of approximately 3400 spatial pixels.

We have used the identifications of \cite{brown2008} and \cite{young2007} to determine the
emission lines of interest. For each of these lines we have fit the line profiles with
Gaussians and extracted the relevant line intensities. These line intensities are given in
Tables~\ref{table:ints1} and \ref{table:ints2}. Additionally, we extracted line
intensities for three \ion{Fe}{9} lines that were recently identified by
\cite{young2009}. This list is not complete. The objective here is to consider the
strongest emission lines that could be used in spatially resolved observations, and not to
consider every emission line in the spectrum.

All of the lines except \ion{Fe}{16} 262.984\,\AA\ are well represented by
Gaussians. There is very little signal in the \ion{Fe}{16} profile and the observed
emission is essentially noise. This measurement does provide a very useful upper bound on
the amount of high temperature emission in this region.

Tables~\ref{table:ints1} and \ref{table:ints2} give the most significant atomic
transitions for each emission line.  The CHIANTI level numbers are also given in these
tables. The level numbers are simply an ordered list of the transitions. They aid in
identifying which emission lines involve transitions to the ground state as well as which
lines originate in the same upper level. The \ion{Fe}{10} 190.038 and 184.536\,\AA\ lines,
for example, originate in the same upper level and form a branching ratio. The branching
ratio only depends on the relative decay rates and should be more accurate than other
ratios. In these tables the emissivity at the peak of the ionization fraction and a density
of $\log n_e = 8.35$ is also given.

In addition to the Fe lines EIS also observes several weaker Si\,\textsc{vii},
\textsc{ix}, and \textsc{x} emission lines. Si emission lines from these ionization stages
have been used in previous emission measure analysis of the quiet corona (e.g.,
\citealt{feldman1999a,warren1999c,landi2002,warren2002a}) and provide a useful comparison
for the analysis of the Fe lines.

Elemental abundances play an important role in determining the magnitude of radiative
losses in the corona. There are several emission lines from high first ionization
potential elements observed within the EIS wavelength ranges \citep{feldman2008}. At
coronal temperatures the \ion{S}{10} 264.233\,\AA\ and \ion{S}{13} 256.686\,\AA\ lines can
be used for studying the composition. Only the \ion{S}{10} line appears in these data. The
\ion{O}{6} lines at 183.937 and 184.117\,\AA\ provide another measurement from a high FIP
element. \ion{O}{6} is Li-like and the ionization fraction for this ion is significant at
coronal temperatures.

\section{Temperature and Density Measurements}

The observed intensities are related to the plasma emissivities, $\epsilon_\lambda(n,T)$,
and the differential emission measure, $\xi(T)$, by the expression
\begin{equation}
  I_\lambda = \frac{1}{4\pi}\int\epsilon_\lambda(n_e,T)\xi(T)\,dT.
  \label{eq:ints}
\end{equation}
In this context the plasma emissivities are the radiated power (erg s$^{-1}$) divided by
the square of the electron number density, $n_e$. For many atomic transitions this
quantity depends mainly on the electron temperature, $T$. For many of the Fe lines
considered here, however, the emissivity is strongly dependent on the density, even for
transitions to the ground state, and to facilitate intensity calculations we have computed
grids of emissivities over a wide range of densities and temperatures using the CHIANTI
5.2.1 atomic physics database (e.g., \citealt{landi2006}). The 5.2.1 version of the
database corrects an error in the atomic data for \ion{Fe}{13}. The abundances of
\cite{feldman1992} and the low density ionization fractions of \cite{mazzotta1998} are
assumed. In this expression the emission measure is the line of sight emission measure,
$\xi(T) = n_e^2\,ds/dT$, and has units of cm$^{-5}$ K$^{-1}$. In the plots we will
generally display the differential emission measure times the temperature.

For this work we will consider three different methods for reconstructing the differential
emission measure from the EIS intensity measurements. The first two methods rely on a
parametrization of the emission measure. The simplest approximation is that of a single
temperature plasma where the differential emission measure is a delta function
\begin{equation}
  \xi(T) = EM_0\,\delta(T-T_0).
\end{equation}
To account for the possibility that there is some dispersion in the temperature
distribution we also consider a Gaussian representation of the differential emission
measure
\begin{equation}
\xi(T) = \frac{EM_0}{\sigma_T\sqrt{2\pi}}
   \exp\left[-\frac{(T-T_0)^2}{2\sigma_T^2}\right]
\end{equation}
The Gaussian DEM is parametrized so that for very narrow temperature distributions we
recover the parameters for the isothermal case, i.e., for $T_0/\sigma_T\gg1$ we have
$\int\xi(T)\,dT\sim EM_0$. To determine the best-fit parameters for either of these
emission measure models we use a Levenberg-Marquardt technique implemented in the
\verb+MPFIT+ package. We have implemented this algorithm so that the density in
Equation~\ref{eq:ints} can either be a free parameter or have a fixed value.

Finally, we also apply a Monte Carlo Markov Chain (MCMC) emission measure algorithm
\citep{kashyap1998,kashyap2000} distributed with the \verb+PINTofALE+ spectral analysis
package to these data. This algorithm has the advantage of not assuming a shape for the
differential emission measure. The MCMC algorithm also provides for estimates of the error
in the DEM. In its current implementation the MCMC algorithm does not allow for the
density to be a free parameter.

\begin{figure}
\centerline{%
\includegraphics[clip,scale=1.0]{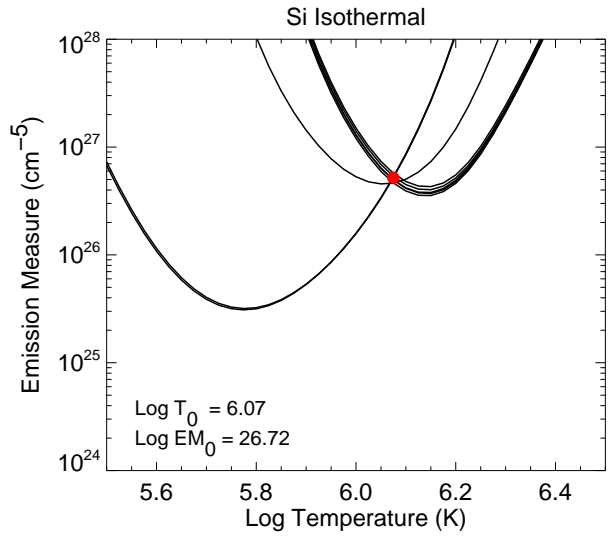}}
\caption{Emission measure analysis of the Si emission lines observed with EIS. In these
  calculations the density is held constant at $\log n_e = 8.35$.}
\label{fig:em_si}
\end{figure}

Perhaps the most important test of the EIS spectra is the application of the isothermal
DEM model to the observed Si emission. The atomic data for Si appears to be very well
understood and previous emission measure analysis in the quiet corona has yielded
consistent results \citep[e.g.,][]{feldman1999a,warren2002a}. The isothermal DEM
calculation for the Si lines is illustrated in Figure~\ref{fig:em_si}, where the best-fit
parameters are shown. Here we also display the emission measure loci curves defined by
\begin{equation}
  \textrm{EM}(T) \equiv \frac{4\pi I_\lambda}{\epsilon_\lambda(n_e,T)}.
\end{equation}
Previous analysis in the quiet corona above the limb has shown that these curves tend to
intersect at a point, suggesting isothermal plasma. As noted before, we solve for the best
fit parameters through $\chi^2$ minimization rather than estimating them by eye as has
been done previously \citep[e.g.,][]{feldman1998}.

The best fit temperature is similar to the values of $\log T_0 = 6.05$ determined by
\cite{landi2002} and $\log T_0 = 6.01$ determined by \cite{allen2000}. Other studies yield
somewhat higher temperatures of $\log T_0 \sim 6.15$
\citep[e.g.,][]{raymond1997,feldman1999a,warren1999c,warren2002a}. Additionally, as
indicated by the values listed in Table~\ref{table:em1}, the computed intensities of the
\ion{Si}{7} and \ion{Si}{10} lines are consistent with each other to within about
$\pm15$\%. This calculation depends on the density, which we have assumed to be $\log n_e
= 8.35$ in this calculation. If we allow the density to be a free parameter in the
minimization then we find $\log n_e = 8.22$. These values are within the range of possible
densities. We will discuss the densities derived from the various line ratios and DEM
inversion methods in detail at the end of this section.

\begin{figure*}
\centerline{%
\includegraphics[clip,scale=1.0]{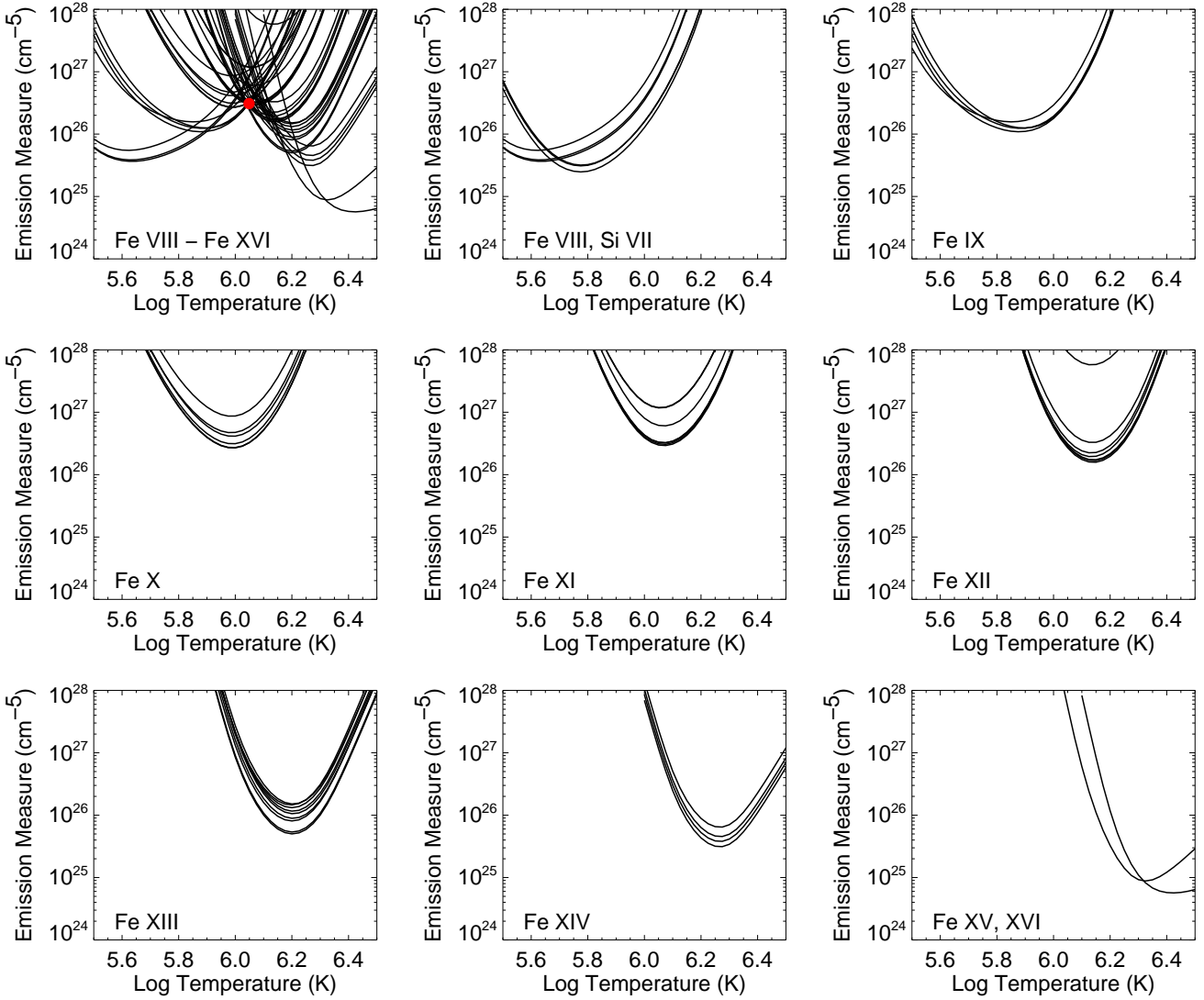}}
\caption{Emission measure loci plots for \ion{Fe}{8}--\textsc{xvi} emission lines observed
  with EIS. These plots illustrate the problems with a number of the emission lines
  observed within this wavelength range. The red dot is the isothermal emission measure
  derived from the Si lines. In these calculations the density is held
  constant at $\log n_e = 8.35$.}
\label{fig:em_loci}
\end{figure*}

The generally good agreement among the observed Si lines is not matched by the observed Fe
spectrum. Emission measure loci plots for all of the Fe lines, which are shown in
Figure~\ref{fig:em_loci}, do not reveal any discernible pattern and do not suggest either
isothermal or multi-thermal plasma. 

If the atomic data, the assumed density, and the observed intensities were mutually
consistent then all of the emission measure loci curves for a given ion would lie very
close together, as they do for \ion{Si}{7} and \ion{Si}{10}. Closer inspection of the the
emission measure loci curves for Fe shown in Figure~\ref{fig:em_loci} suggests that for
each ion there are several emission lines that are mutually consistent and others that are
discrepant by varying amounts. Here we identify these lines and discuss previously
identified blends.

\paragraph{\ion{Fe}{8}} The emission measure loci for 185.213 and 186.601\,\AA\ lines are
consistent, but the curve for 194.663\,\AA\ is about a factor of 2 higher.
\cite{brown2008} indicate a possible blend of \ion{Fe}{8} 194.663\,\AA\ with \ion{O}{5}
195.593\,\AA, which would have negligible intensity at this height above the limb. In the
core of an active region \ion{Fe}{8} 185.213\,\AA\ is blended with \ion{Ni}{16}
185.251\,\AA\ and \ion{Fe}{8} 186.601\,\AA\ is blended with \ion{Ca}{14} 186.610\,\AA. As
stated earlier, the ionization fraction for \ion{Fe}{8} may need revision, so these lines
should be used with caution.

\paragraph{\ion{Fe}{9}} The newly identified 188.497, 189.941, and 197.862\,\AA\ lines are
all in good agreement. The emission measure loci for \ion{Fe}{9} 171.073\,\AA\ is off by
about a factor of 2. The effective area for EIS at this wavelength is very low, making the
observed intensity highly uncertain.

\paragraph{\ion{Fe}{10}} The 174.532, 177.239, and 184.536\,\AA\ lines are all in
agreement. The \ion{Fe}{10} 190.038, 207.449, and 257.262 lines are not. The 184.536 and
190.038\,\AA\ lines originate in the same upper level and form a branching ratio. The
theoretical ratio, however, is 3.56 while the observed ratio is 2.70. The \ion{Fe}{10}
257.262 line forms a density sensitive ratio with 184.536 and 190.038\,\AA, but the
problems with the emission measure loci suggests that the densities derived from this
ratio are not consistent with the densities derived from the Si ratio. \cite{brown2008}
indicate a possible blend of \ion{Fe}{10} 184.536\,\AA\ with \ion{Ar}{11} 184.524\,\AA,
which would be a problem for active region observations.

\paragraph{\ion{Fe}{11}} The 180.401, 182.167, 188.216, and 192.813\,\AA\ lines are all in
agreement. The emission measure loci for 188.299, 257.547, and 257.772\,\AA\ differ by
factors of 2 to 4. The 182.167 and 188.216\,\AA\ lines form a density sensitive line
ratio. \cite{brown2008} indicate that \ion{Fe}{11} 180.401\,\AA\ is blended with
\ion{Fe}{10} 180.407\,\AA. Using the isothermal emission measure determined from Si to
estimate the intensities of the two lines suggests that the \ion{Fe}{10} contribution to
the observed intensity is about 10\%. There are \ion{O}{5} lines that can make important
contributions to the observed \ion{Fe}{11} 192.813\,\AA\ line profile during transient
events \citep{ko2009}.

\paragraph{\ion{Fe}{12}} The \ion{Fe}{12} 186.880, 192.394, 193.509, 195.119, and
196.640\,\AA\ lines are all in relatively good agreement. The 203.720 and 256.925\,\AA\
lines differ by factors of 2 and 30, respectively. The 186.880 and 196.640\,\AA\ lines
form density sensitive ratios when paired with any of the 192.394, 193.509, and
195.119\,\AA\ lines. There is a \ion{S}{11} line at 186.84\,\AA, but it is generally weak
compared with the \ion{Fe}{12} 186.880 line \citep{young2008}.  In active regions,
\cite{young2008} note that there is an \ion{Fe}{12} 195.18\,\AA\ line that becomes
important at high densities. This component is generally small but can impact Doppler
shift and line width measurements with the \ion{Fe}{12} 195.119\,\AA\ line.

\paragraph{\ion{Fe}{13}} The 196.525, 197.434, 200.021, 202.044, and 203.826\,\AA\ lines
are generally consistent. The emission measure loci for 201.121, 204.937, 246.208, and
251.953 are all offset by varying amounts. The 196.525 and 203.826\,\AA\ lines form
density sensitive ratios with 202.044\,\AA. There is a possible blend of \ion{Fe}{13}
196.525\,\AA\ with \ion{Fe}{8} 196.65\,\AA, but the \ion{Fe}{8} line is believed to be
weak \citep{young2008}.  \cite{brown2008} indicate that \ion{Fe}{13} 201.121\,\AA\ is
blended with \ion{Fe}{12} 201.121\,\AA.

\vspace{6pt}

\paragraph{\ion{Fe}{14}} The 211.316, 270.519, and 274.203\,\AA\ emission measure loci
curves are all relatively close to each other while the curve for 264.787\,\AA\ is
offset. The 264.787 and 274.203\,\AA\ lines form a density sensitive pair, but this ratio
should be in the low density limit in the quiet Sun. There is a blend of \ion{Fe}{14}
274.203\,\AA\ with \ion{Si}{7} 274.175\,\AA. \cite{young2007} indicate that the
\ion{Si}{7} 274.175\,\AA\ intensity is less than 0.25 times the intensity of \ion{Si}{7}
275.352\,\AA. The ratio of the Si lines is sensitive to density, however, and for the the
low densities measured above the limb the ratio is calculated to be about 10\%, suggesting
a negligible contribution of \ion{Si}{7} to the \ion{Fe}{14} 274.203\,\AA\ intensity
measured here.

\paragraph{\ion{Fe}{15} and \ion{Fe}{16}} There is only a single \ion{Fe}{15} emission
line in the EIS wavelength ranges. There are several \ion{Fe}{16} lines present, but as
indicated earlier, no emission is observed in any of these lines. The intensity given in
Table~\ref{table:ints1} for \ion{Fe}{16} 262.984\,\AA\ serves only as an upper bound and
limits the peak temperature in the emission measure. There is no independent verification
of the consistency of these lines.  

\begin{figure*}[t!]
\centerline{%
\includegraphics[clip,scale=1.0]{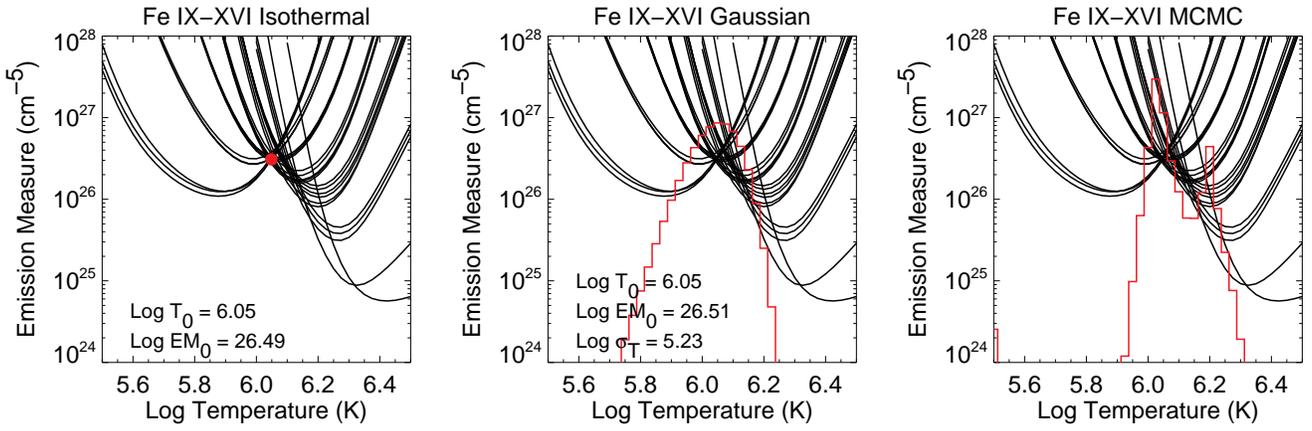}}
\caption{Differential emission measures derived from Fe\,\textsc{ix}--\textsc{xvi} and
  different emission measure models (isothermal, Gaussian, and Monte Carlo Markov
  chain). For the Gaussian and MCMC models the temperature times the differential emission
  measure is displayed. The MCMC DEM has the lowest $\chi^2$ and best reproduces the
  observed intensities. In these calculations the density is held constant at $\log n_e =
  8.35$.}
\label{fig:em_fe}
\end{figure*}

We have used the lines identified here by their internal consistency to compute the
emission measure using the three different models. Both the emission measures and the
emission measure loci are shown in Figure~\ref{fig:em_fe}. The isothermal solution for
these selected Fe lines is very similar the solution found for the Si lines. The best fit
temperature is $\log T = 6.05$, compared with 6.07 for the Si lines. It is clear, however,
that the isothermal model cannot reproduce both the low temperature emission
(\ion{Fe}{9}--\textsc{xiii}) and the emission observed in the higher temperature lines
(\ion{Fe}{14}--\textsc{xvi}). The intensities for the high temperatures lines computed
from the isothermal emission measure are systematically too small by a factor of about 4.

This systematic discrepancy suggests that the plasma in the quiet corona is not
isothermal. To investigate this we have applied the Gaussian DEM algorithm to the Fe
lines. If we include only the \ion{Fe}{9}-\textsc{xiii} lines we obtain a very narrow
emission measure distribution ($\sigma_T=4.8$).  Including the \ion{Fe}{14}--\textsc{xvi}
lines leads to a somewhat broader emission measure distribution ($\sigma_T=5.2$), but only
a marginal improvement between the observed and calculated intensities. For this case the
observed intensities are systematically larger than what is calculated from the Gaussian
emission measure by about a factor of 2.

The failure of the Gaussian emission measure model to reproduce all of the observed line
emission is due in large part to the symmetry that this model imposes on the
DEM. Increases in the emission measure at higher temperature must be accompanied by
increases at lower temperatures, and this would lead to discrepancies in the \ion{Fe}{9}
and \ion{Fe}{10} intensities. The MCMC algorithm does not assume any functional form for
the emission measure and provides for more flexibility. The differential emission measure
calculated using the MCMC algorithm is displayed in Figure~\ref{fig:em_fe} and calculated
intensities are given in Table~\ref{table:em2}. In this case the agreement between the
observed and calculated intensities is improved for the high temperature lines while
maintaining the good agreement at the lower temperatures. This is achieved by the
introduction of a high temperature tail in the differential emission measure in addition
to the strong peak near 1\,MK. The strong peak in the emission measure is consistent with
previous work. Previous studies have usually employed relatively restrictive
parametrizations for the emission measure and would not have been able to identify this
high temperature component. 

The best fit parameters for the Si and Fe isothermal emission measure calculations show a
significant discrepancy in the magnitude of the emission measure, with the Si emission
measure being about 60\% larger ($\log EM_0 = 26.49$ for Fe and 26.72 for Si). We find a
similar difference if we use the MCMC DEM to infer the intensities of the Si
lines. Curiously, the Si lines all fall on the long wavelength detector while the Fe lines
all fall on the short wavelength detector. None of the long wavelength \ion{Fe}{10},
\textsc{xi}, \textsc{xii}, or \textsc{xiii} lines are consistent with the short wavelength
lines from the same ion. Similarly, the high temperature lines, whose intensities the
isothermal models systematically under-predict, generally lie on the long wavelength
detector. This suggests that calibration differences between the two detectors might
explain at least part of these discrepancies. While this explanation may sound plausible
it does not withstand closer scrutiny. Perhaps most importantly, the magnitude of the
discrepancy between the calculated and observed \ion{Fe}{14}--\textsc{xvi} intensities is
much larger than the difference between the Si and Fe isothermal emission measures. If the
observed values were reduced by a factor of 1.6 the calculated intensities would still be
off by a factor of 3 or more. One \ion{Fe}{14} line does appear on the short wavelength
detector (211.316\,\AA) and its behavior is similar to the long wavelength \ion{Fe}{14}
lines. Finally, the discrepancies in the emission measure loci curves discussed previously
do not show any systematic variation with wavelength. With these data we find no evidence
that there is a problem with the relative calibration of the detectors.

\begin{figure*}[t!]
\centerline{%
\includegraphics[clip,scale=1.0]{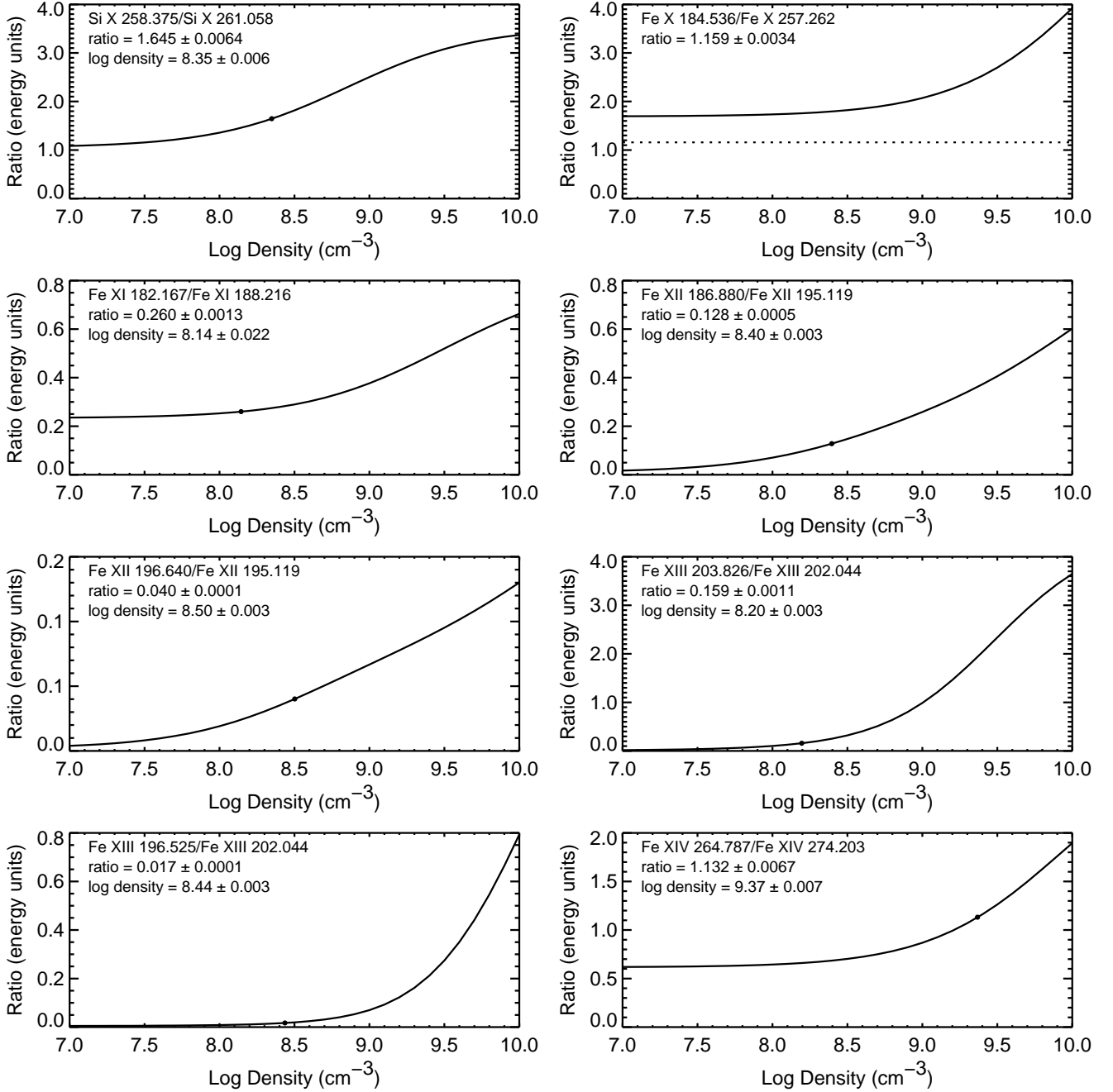}}
\caption{Electron densities derived from three density sensitive line ratios. The
  emissivities were evaluated at $\log T = 6.05$.}
\label{fig:density}
\end{figure*}

When computing the line intensity the density and temperature both play an important role
in determining the emissivity. This makes it difficult to determine the temperature and
density independently. One way of addressing this issue is to allow both parameters to
vary while using Equation~\ref{eq:ints} to find the optimal parameters. We have done this
in applying the isothermal model to the Si and \ion{Fe}{9}-\textsc{xiii} lines and we
generally find somewhat different densities. The EIS wavelength ranges contain a large
number of density sensitive line ratios (see \citealt{young2007} for a summary) and we can
use these data to compare the results. In Figure~\ref{fig:density} we show the densities
derived directly from \ion{Si}{10} 258.375/261.058\,\AA, \ion{Fe}{10}
184.536/257.262\,\AA, \ion{Fe}{11} 182.167/188.216\,\AA, \ion{Fe}{12}
186.880/195.119\,\AA, \ion{Fe}{12} 196.640/195.119\,\AA, \ion{Fe}{13}
196.525/202.044\,\AA, \ion{Fe}{13} 203.826/202.044\,\AA, and \ion{Fe}{14}
264.787/274.203\,\AA.  The emissivities are computed assuming $\log T = 6.05$, as derived
from the isothermal emission measure analysis.

These ratios show some significant differences. The \ion{Fe}{10} and \ion{Fe}{14} ratios
give useless results, with the \ion{Fe}{10} being below the low density limit and
\ion{Fe}{14} yielding a density an order of magnitude larger than the others. The other
ratios yield values ranging from $\log n_e = 8.14$ to 8.50. These values are generally
consistent with the densities determined in earlier analysis of quiet Sun limb
observations, e.g., $\log n_e\sim 8.3$ by \citealt{doschek1997b} and $\log n_e\sim 8.5$ by
\citealt{landi2002}. Based on previous analysis, the \ion{Si}{10} ratio is probably the
most reliable ratio of this group. We note that the densities from the Fe lines are
scattered around the density derived from Si and if we average the Fe densities together
we obtain $8.34\pm0.16$. This is a variation of about a factor of 2. To simplify the
discussion, all of the plots and tables assume a density of $\log n_e = 8.35$.

The large region that we have used for computing an average spectrum leads to very small
statistical errors for the measured intensities. \cite{lang2006} calculate the uncertainty
in the absolute radiometric calibration for EIS to be 22\%. This additional uncertainty
can be easily added in quadrature to the values given in Tables~\ref{table:ints1} and
\ref{table:ints2}. We have rerun all of the dem calculations including the calibration
uncertainties and obtained very similar results. The biggest differences are for the MCMC
algorithm, which yields a somewhat smoother emission measure at high
temperatures. Variations in the detailed structure of the emission measure are not
surprising given the ill-posed nature of the emission measure inversion problem
\citep[e.g.,][]{craig1976}.

\section{Abundances}

To accurately compute the magnitude of the radiative losses in the corona we must have
some measure of the coronal composition. Understanding radiative cooling is important for
modeling the evolution of coronal loops \cite[e.g.,][]{warren2003,winebarger2004}.
Spatially averaged measurements suggest that the low first ionization potential elements,
such as Fe, Si, and Mg, are all enriched in the corona relative to high FIP elements, such
as C, N, and O by about a factor of 4 \citep[e.g.,][]{feldman1998}. Temporally resolved
measurements in active regions indicate, however, that this enhancement may change with
time \citep{widing2001}, with newly emerged active regions having more photospheric
abundances (i.e., no enrichment of the low FIP elements). This suggests the need for more
systematic abundance measurements.

Several abundance diagnostics in the EIS wavelength ranges have been discussed by
\cite{feldman2008}. For coronal plasma the most useful EIS emission lines from high FIP
elements are \ion{S}{10} 264.233\,\AA\ and \ion{S}{13} 256.686\,\AA. In these quiet Sun
data only the \ion{S}{10} 264.233\,\AA\ line appears. The FIP for S is 10.3\,eV and it
lies at the boundary between the high and low FIP elements. Some measurements
\citep[e.g.,][]{feldman1998} show a partial enrichment of S in the corona. Other
measurements \citep[e.g.,][]{feldman1992} indicate essentially photospheric
abundances. The general trend is for S to behave somewhat differently than the low FIP
elements, and \ion{S}{10} can be used as a proxy for measuring relative abundance
variations in EIS observations of the corona. For these observations above the quiet
corona it is also possible to use \ion{O}{6} as a high FIP line
\citep[e.g.,][]{feldman1998}. \ion{O}{6} is Li-like and the ionization fraction extends to
high temperatures.

In calculating the emissivities we have assumed a coronal composition with $A_{Fe} =
8.10$, $A_{Si} = 8.10$, $A_{O} = 8.89$, and $A_{S} = 7.27$ \citep{feldman1992}. The
photospheric composition of \cite{grevesse1998} has $A_{Fe} = 7.50$, $A_{Si} = 7.55$,
$A_{O} = 8.83$, and $A_{S} = 7.33$.  If we use the MCMC emission measure computed from the
Fe lines to compute the the intensities of \ion{O}{6} 183.937, 184.117 and \ion{S}{10}
264.233\,\AA\ we obtain values of 0.8, 1.6 and 15.0\,erg~cm$^{-2}$~s$^{-1}$~sr$^{-1}$
respectively. The observed intensities in this region are 2.8, 4.8 and
34.8\,erg~cm$^{-2}$~s$^{-1}$~sr$^{-1}$, suggesting that the relative abundances of the
\citep{feldman1992} are not consistent with these data. The abundance of S and O in the
corona would need to be increased by about a factor of 2.9 or the abundance of Fe
decreased by about a factor of 2.9 to resolve this discrepancy. Similarly, the difference
in the emission measures derived from the Si and Fe lines could be resolved by adjusting
their relative abundances.

\section{Quiet Sun Disk}

The observations taken above the limb provide an opportunity to study a high signal to
noise spectrum derived from an average over a largely homogeneous region. This analysis
has revealed many problems with interpreting the observed intensities. These problems
could be due to blends with other lines, errors in the calibration, or blends with other
lines. Since the morphology of the quiet solar atmosphere changes with temperature
\cite[e.g.,][]{feldman1999b} examining spatially resolved images in these emission lines
offers a means of identifying previously unidentified blends in the lines.

Most EIS observations preserve the information from only a few narrow spectral
windows. Observations that contain the full wavelength range are typically small in
size. We have found several observations that contain the full wavelength range and cover
a relatively large region on the Sun (up to $128\arcsec\times512\arcsec$). For one of
these observations (taken November 15, 2007 11:14 UT) we have processed the data and
constructed rasters in as many of the emission lines given in Tables~\ref{table:ints1} and
\ref{table:ints2} as possible. These rasters are computed by performing Gaussian fits to
the spectral lines at each spatial position. Unfortunately, some emission lines from the
limb spectra are too weak to fit in the spatially resolved disk spectra.

The rasters for many of the emission lines of interest are shown in
Figures~\ref{fig:disk1} and \ref{fig:disk2}. These rasters suggest a progression from
small scale structures that correspond to the magnetic network at low temperatures to much
longer loops, with no network pattern at high temperatures. Surprisingly, there is little
evidence for blending in many of the problem emission lines. For example, the \ion{Fe}{10}
184.536 and 190.038\,\AA\ rasters are nearly identical. The contrast in the \ion{Fe}{10}
257.262\,\AA\ raster is somewhat different than in the other rasters, but this line is
sensitive to density and this is to be expected. Similarly, the \ion{Fe}{11} 188.216 and
188.299\,\AA\ rasters are nearly identical. There is some evidence for small bright
features in the \ion{Fe}{11} 192.813\,\AA\ raster, suggestive of \ion{O}{5} emission. For
\ion{Fe}{12} the discrepant 203.720\,\AA\ rasters appears to be consistent with the
rasters from the other lines. The \ion{Fe}{12} 256.925\,\AA\ line does appear to be
blended with a cooler line. The contrast between the center of this raster and the
structures in the north and south is smaller than for the other \ion{Fe}{12} lines. For
similar reasons, the \ion{Fe}{14} 264.787\,\AA\ also appears to be blended with a cooler
line.

\section{Discussion}

We have presented a detailed analysis of spectroscopic observations taken in the quiet
corona above the limb with the EIS instrument on \textit{Hinode}. The strongly peaked
differential emission measure calculated from the Si and Fe lines are generally consistent
with each other and with previous measurements. This suggests that despite the complexity
of the Fe atom, accurate differential emission measure calculations are possible with the
strongest emission lines observed with EIS. Our analysis validates some of the initial EIS
emission measure calculations for coronal loops,
\citep[e.g.,][]{warren2008b,tripathi2009}, which emphasize the lines given in
Table~\ref{table:em2}.

Using the MCMC DEM algorithm we find evidence for a tail in the emission measure that
extends to high temperatures. Such features are extremely important to theories of coronal
heating based on the impulsive release of energy
\citep[e.g.][]{cargill2004,patsourakos2006,patsourakos2008}. It will be interesting to see
if similar features of the DEM are observed in active region loops. 

The densities measured from 6 of the 8 density sensitive line ratios that we considered
are generally consistent with each other and with previous measurements. There is,
however, considerable scatter in the measured densities.  Measurements in active regions
suggest that these differences are systematic, with the \ion{Fe}{12} 186.880/195.119\,\AA\
density being systematically higher than the \ion{Fe}{13} 203.826/202.044\,\AA\ density
\citep{young2007}.

This work has highlighted problems with the analysis of many emission lines in the EIS
spectral wavelengths. These differences result from some combination of errors in the
atomic data, blends with unidentified lines, and uncertainties with the EIS
calibration. Some of these difficulties are clearly due to the errors in the atomic data
or assumptions that have been made in using them. The observations suggest that the
ionization fractions for \ion{Fe}{8}, for example, may need to be shifted to high
temperatures. The quiet Sun disk rasters for \ion{Fe}{10} 190.038\,\AA\ and 257.262\,\AA,
\ion{Fe}{11} 188.299\,\AA, \ion{Fe}{12} 203.720\,\AA, and \ion{Fe}{13} 251.953\,\AA\ do
not appear to be blended with other lines. These rasters do suggest blends for
\ion{Fe}{12} 256.925\,\AA\ and \ion{Fe}{14} 264.787\,\AA. The possibility of errors in the
EIS relative calibration is unclear. The \ion{Fe}{14} 211.316\,\AA\ line is consistent
with \ion{Fe}{14} 270.519 and 274.203\,\AA, suggesting that there are no significant
discrepancies.  It is unsettling, however, that none of the \ion{Fe}{10}, \textsc{xi},
\textsc{xii}, and \textsc{xiii} lines near 250\,\AA\ agree with the lines at shorter
wavelengths.

\begin{figure*}[t!]
\centerline{%
\includegraphics[clip,scale=0.85]{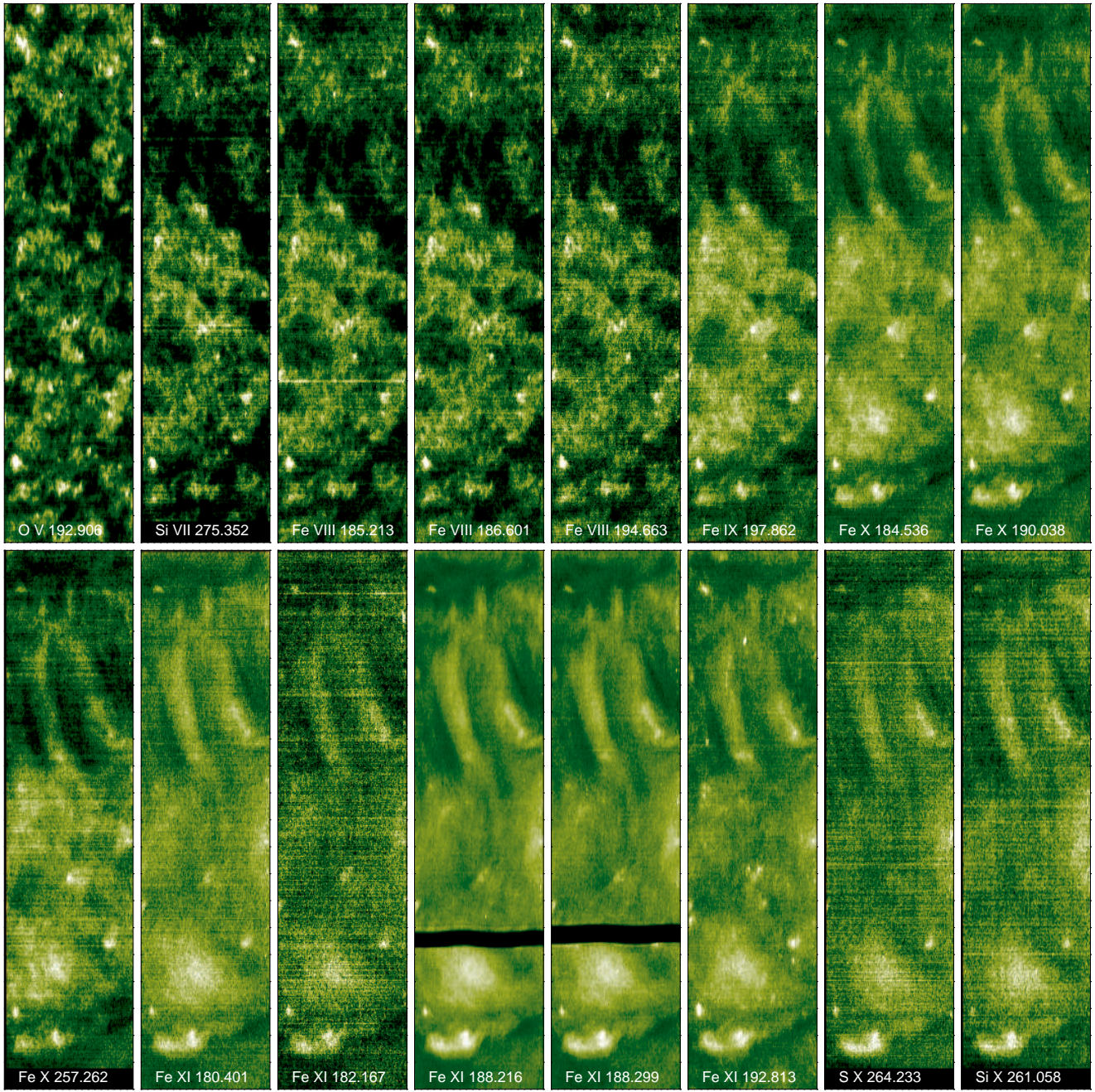}}
\caption{EIS rasters from observations of the quiet solar disk taken November 15, 2007
  from 11:13:59 to 14:12:17 UT. The 1\arcsec\ slit has been stepped over a region
  $128\arcsec\times512\arcsec$ in size, and an 80\,s exposure has been taken at each slit
  position. Rasters for \ion{O}{5} -- \ion{Si}{10} are shown. These spatially resolved
  disk measurements allow for the morphology of different emission lines from the same ion
  to be compared. The \ion{Fe}{11} 192.813\,\AA\ shows a small contribution from
  \ion{O}{5}.}
\label{fig:disk1}
\end{figure*}

\begin{figure*}[t!]
\centerline{%
\includegraphics[clip,scale=0.85]{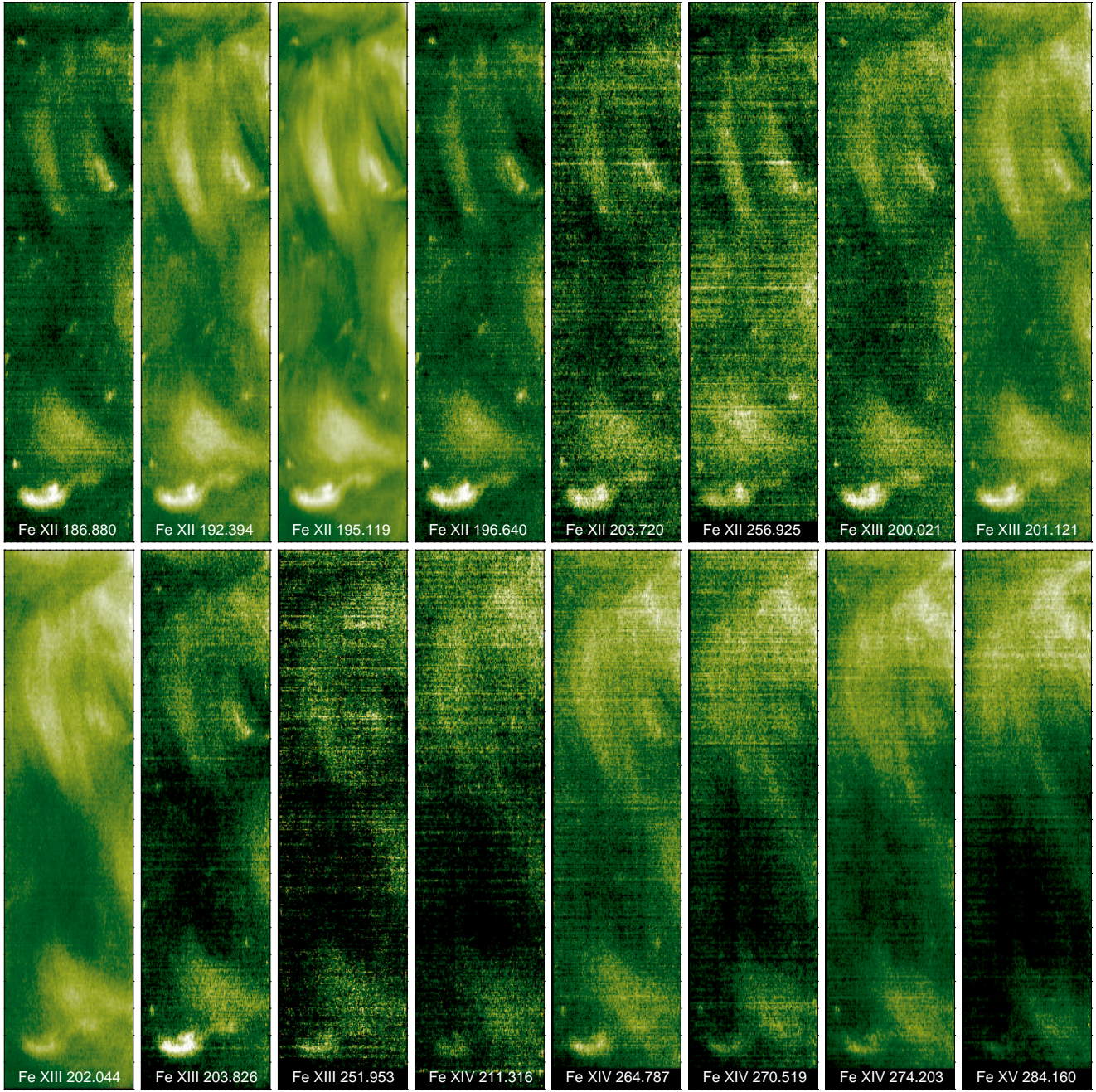}}
\caption{The same region as is shown in Figure~\protect{\ref{fig:disk1}} except that
  \ion{Fe}{12} -- \ion{Fe}{15} are shown. The \ion{Fe}{12} 256.925\,\AA\ and \ion{Fe}{14}
  264.787\,\AA\ lines appear to have excess emission in the center of the rasters,
  suggesting blends with cooler emission lines.}
\label{fig:disk2}
\end{figure*}

The use of EIS observations for abundance measurements is still unsettled. The ratios of
the S and O lines to Si and Fe do not yield results that are consistent with existing sets
of solar abundances. It is unclear how these issues can be resolved at present. Despite
this difficulty it is still possible to use EIS observations to investigate spatial and
temporal variations in the solar composition. This is illustrated in
Figure~\ref{fig:disk1}, where we can see spatially resolved images in \ion{S}{10} and
\ion{Si}{10}. When combined with emission measure analysis, observations such as these can
be used to understand the how much the composition varies from structure to structure in
the solar corona. Work on the analysis of quiet Sun disk spectra are currently in progress
and will be reported in a future paper \cite{brooks2009}.


\acknowledgments Hinode is a Japanese mission developed and launched by ISAS/JAXA, with
NAOJ as domestic partner and NASA and STFC (UK) as international partners. It is operated
by these agencies in co-operation with ESA and NSC (Norway).



\clearpage

\clearpage

\clearpage

\clearpage

\clearpage

\clearpage

\clearpage

\begin{deluxetable}{llcr@{\quad---\quad}lrrcr@{\quad$\pm$\quad}rlc}
\tabletypesize{\scriptsize}
\tablewidth{0mm}
\tablecaption{Fe Intensities Measured in the Quiet Corona above the Limb with EIS\tablenotemark{a}}
\tablehead{
\multicolumn{1}{c}{Ion} &
\multicolumn{1}{c}{$\lambda$} &
\multicolumn{1}{c}{$T_{max}$} &
\multicolumn{2}{c}{Transition} &
\multicolumn{1}{c}{L1} &
\multicolumn{1}{c}{L2} &
\multicolumn{1}{c}{$\epsilon_\lambda$} &
\multicolumn{1}{c}{$I_{obs}$} &
\multicolumn{1}{c}{$\sigma_{Iobs}$} &
\multicolumn{1}{c}{}
}
\startdata
\ion{Fe}{8} & 185.213 & 5.57 & 3p$^6$ 3d $^2$D$_{5/2}$ & 3p$^5$ 3d$^2$ ($^3$F) $^2$F$_{7/2}$ & 2 & 46 & 6.23e-24 & 18.54 & 0.11 & \\
\ion{Fe}{8} & 186.601 & 5.57 & 3p$^6$ 3d $^2$D$_{3/2}$ & 3p$^5$ 3d$^2$ ($^3$F) $^2$F$_{5/2}$ & 1 & 45 & 4.63e-24 & 14.60 & 0.09 & \\
\ion{Fe}{8} & 194.663 & 5.57 & 3p$^6$ 3d $^2$D$_{5/2}$ & 3p$^6$ 4p $^2$P$_{3/2}$ & 2 & 43 & 1.22e-24 & 5.38 & 0.03 & \\
\ion{Fe}{9} & 171.073 & 5.81 & 3s$^2$ 3p$^6$ $^1$S$_{0}$ & 3s$^2$ 3p$^5$ 3d $^1$P$_{1}$ & 1 & 13 & 6.96e-23 & 921.27 & 17.93 & \\
\ion{Fe}{9} & 188.497 & 5.81 & 3s$^2$ 3p$^5$ 3d $^3$F$_{4}$ & 3s$^2$ 3p$^4$ ($^3$P) 3d$^2$ $^3$G$_{5}$ & 5 & 95 & 2.66e-24 & 31.28 & 0.12 & $\heartsuit$ \\
\ion{Fe}{9} & 189.941 & 5.81 & 3s$^2$ 3p$^5$ 3d $^3$F$_{3}$ & 3s$^2$ 3p$^4$ ($^3$P) 3d$^2$ $^3$G$_{4}$ & 6 & 94 & 1.55e-24 & 15.36 & 0.05 & $\heartsuit$ \\
\ion{Fe}{9} & 197.862 & 5.81 & 3s$^2$ 3p$^5$ 3d $^1$P$_{1}$ & 3s$^2$ 3p$^5$ 4p $^1$S$_{0}$ & 13 & 140 & 1.65e-24 & 21.02 & 0.06 & $\heartsuit$ \\
\ion{Fe}{10} & 174.532 & 5.99 & 3s$^2$ 3p$^5$ $^2$P$_{3/2}$ & 3s$^2$ 3p$^4$ ($^3$P) 3d $^2$D$_{5/2}$ & 1 & 30 & 2.64e-23 & 572.57 & 4.44 & $\heartsuit$ \\
\ion{Fe}{10} & 177.239 & 5.99 & 3s$^2$ 3p$^5$ $^2$P$_{3/2}$ & 3s$^2$ 3p$^4$ ($^3$P) 3d $^2$P$_{3/2}$ & 1 & 28 & 1.44e-23 & 308.28 & 1.75 & $\heartsuit$ \\
\ion{Fe}{10} & 184.536 & 5.99 & 3s$^2$ 3p$^5$ $^2$P$_{3/2}$ & 3s$^2$ 3p$^4$ ($^1$D) 3d $^2$S$_{1/2}$ & 1 & 27 & 5.68e-24 & 142.17 & 0.29 & $\heartsuit$ \\
\ion{Fe}{10} & 190.038 & 5.99 & 3s$^2$ 3p$^5$ $^2$P$_{1/2}$ & 3s$^2$ 3p$^4$ ($^1$D) 3d $^2$S$_{1/2}$ & 2 & 27 & 1.60e-24 & 52.74 & 0.11 & \\
\ion{Fe}{10} & 207.449 & 5.99 & 3s$^2$ 3p$^5$ $^2$P$_{3/2}$ & 3s$^2$ 3p$^4$ ($^1$D) 3d $^2$F$_{5/2}$ & 1 & 23 & 3.49e-25 & 24.09 & 0.25 & \\
\ion{Fe}{10} & 257.262 & 5.99 & 3s$^2$ 3p$^5$ $^2$P$_{3/2}$ & 3s$^2$ 3p$^4$ ($^3$P) 3d $^4$D$_{7/2}$ & 1 & 5 & 2.57e-24 & 122.67 & 0.25 & \\
& & & 3s$^2$ 3p$^5$ $^2$P$_{3/2}$ & 3s$^2$ 3p$^4$ ($^3$P) 3d $^4$D$_{5/2}$ & 1 & 4 & 6.51e-25 & & & \\
\ion{Fe}{11} & 180.401 & 6.07 & 3s$^2$ 3p$^4$ $^3$P$_{2}$ & 3s$^2$ 3p$^3$ ($^4$S) 3d $^3$D$_{3}$ & 1 & 42 & 1.75e-23 & 432.24 & 1.00 & $\heartsuit$ \\
\ion{Fe}{11} & 182.167 & 6.07 & 3s$^2$ 3p$^4$ $^3$P$_{1}$ & 3s$^2$ 3p$^3$ ($^4$S) 3d $^3$D$_{2}$ & 2 & 43 & 2.29e-24 & 58.50 & 0.28 & $\heartsuit$ \\
\ion{Fe}{11} & 188.216 & 6.07 & 3s$^2$ 3p$^4$ $^3$P$_{2}$ & 3s$^2$ 3p$^3$ ($^2$D) 3d $^3$P$_{2}$ & 1 & 38 & 8.20e-24 & 224.90 & 0.25 & $\heartsuit$ \\
\ion{Fe}{11} & 188.299 & 6.07 & 3s$^2$ 3p$^4$ $^3$P$_{2}$ & 3s$^2$ 3p$^3$ ($^2$D) 3d $^1$P$_{1}$ & 1 & 39 & 2.98e-24 & 153.09 & 0.18 & \\
\ion{Fe}{11} & 192.813 & 6.07 & 3s$^2$ 3p$^4$ $^3$P$_{1}$ & 3s$^2$ 3p$^3$ ($^2$D) 3d $^3$P$_{2}$ & 2 & 38 & 1.71e-24 & 57.73 & 0.12 & $\heartsuit$ \\
& & & 3s$^2$ 3p$^4$ $^3$P$_{1}$ & 3s$^2$ 3p$^3$ ($^2$D) 3d $^3$S$_{1}$ & 2 & 37 & 3.86e-25 & & & \\
\ion{Fe}{11} & 257.547 & 6.07 & 3s$^2$ 3p$^4$ $^3$P$_{2}$ & 3s$^2$ 3p$^3$ ($^4$S) 3d $^5$D$_{3}$ & 1 & 13 & 2.33e-25 & 24.29 & 0.13 & \\
\ion{Fe}{11} & 257.772 & 6.07 & 3s$^2$ 3p$^4$ $^3$P$_{2}$ & 3s$^2$ 3p$^3$ ($^4$S) 3d $^5$D$_{2}$ & 1 & 12 & 1.14e-25 & 11.72 & 0.07 & \\
\ion{Fe}{12} & 186.880 & 6.13 & 3s$^2$ 3p$^3$ $^2$D$_{5/2}$ & 3s$^2$ 3p$^2$ ($^3$P) 3d $^2$F$_{7/2}$ & 3 & 39 & 1.90e-24 & 35.17 & 0.13 & $\heartsuit$ \\
& & & 3s$^2$ 3p$^3$ $^2$D$_{3/2}$ & 3s$^2$ 3p$^2$ ($^3$P) 3d $^2$F$_{5/2}$ & 2 & 36 & 3.40e-25 & & & \\
\ion{Fe}{12} & 192.394 & 6.13 & 3s$^2$ 3p$^3$ $^4$S$_{3/2}$ & 3s$^2$ 3p$^2$ ($^3$P) 3d $^4$P$_{1/2}$ & 1 & 30 & 5.67e-24 & 79.48 & 0.12 & $\heartsuit$ \\
\ion{Fe}{12} & 193.509 & 6.13 & 3s$^2$ 3p$^3$ $^4$S$_{3/2}$ & 3s$^2$ 3p$^2$ ($^3$P) 3d $^4$P$_{3/2}$ & 1 & 29 & 1.19e-23 & 177.53 & 0.16 & $\heartsuit$ \\
\ion{Fe}{12} & 195.119 & 6.13 & 3s$^2$ 3p$^3$ $^4$S$_{3/2}$ & 3s$^2$ 3p$^2$ ($^3$P) 3d $^4$P$_{5/2}$ & 1 & 27 & 1.77e-23 & 274.67 & 0.17 & $\heartsuit$ \\
\ion{Fe}{12} & 196.640 & 6.13 & 3s$^2$ 3p$^3$ $^2$D$_{5/2}$ & 3s$^2$ 3p$^2$ ($^1$D) 3d $^2$D$_{5/2}$ & 3 & 34 & 6.04e-25 & 11.03 & 0.04 & $\heartsuit$ \\
\ion{Fe}{12} & 203.720 & 6.13 & 3s$^2$ 3p$^3$ $^2$D$_{5/2}$ & 3s$^2$ 3p$^2$ ($^1$S) 3d $^2$D$_{5/2}$ & 3 & 32 & 7.35e-25 & 20.32 & 0.12 & \\
\ion{Fe}{12} & 256.925 & 6.13 & 3s$^2$ 3p$^3$ $^2$D$_{3/2}$ & 3s$^2$ 3p$^2$ ($^3$P) 3d $^4$F$_{5/2}$ & 2 & 15 & 6.77e-26 & 39.09 & 0.15 & \\
& & & 3s 3p$^4$ $^4$P$_{5/2}$ & 3s 3p$^3$ 3d $^4$D$_{7/2}$ & 6 & 50 & 1.08e-26 & & & \\
\ion{Fe}{13} & 196.525 & 6.19 & 3s$^2$ 3p$^2$ $^1$D$_{2}$ & 3s$^2$ 3p 3d $^1$F$_{3}$ & 4 & 26 & 2.48e-25 & 2.71 & 0.02 & $\heartsuit$ \\
\ion{Fe}{13} & 197.434 & 6.19 & 3s$^2$ 3p$^2$ $^3$P$_{0}$ & 3s$^2$ 3p 3d $^3$D$_{1}$ & 1 & 23 & 7.54e-25 & 7.10 & 0.04 & $\heartsuit$ \\
\ion{Fe}{13} & 200.021 & 6.19 & 3s$^2$ 3p$^2$ $^3$P$_{1}$ & 3s$^2$ 3p 3d $^3$D$_{2}$ & 2 & 25 & 9.70e-25 & 9.43 & 0.06 & $\heartsuit$ \\
\ion{Fe}{13} & 201.121 & 6.19 & 3s$^2$ 3p$^2$ $^3$P$_{1}$ & 3s$^2$ 3p 3d $^3$D$_{1}$ & 2 & 23 & 3.68e-24 & 45.14 & 0.12 & \\
\ion{Fe}{13} & 202.044 & 6.19 & 3s$^2$ 3p$^2$ $^3$P$_{0}$ & 3s$^2$ 3p 3d $^3$P$_{1}$ & 1 & 20 & 1.82e-23 & 157.66 & 0.26 & $\heartsuit$ \\
\ion{Fe}{13} & 203.826 & 6.19 & 3s$^2$ 3p$^2$ $^3$P$_{2}$ & 3s$^2$ 3p 3d $^3$D$_{3}$ & 3 & 24 & 2.86e-24 & 25.02 & 0.16 & $\heartsuit$ \\
& & & 3s$^2$ 3p$^2$ $^3$P$_{2}$ & 3s$^2$ 3p 3d $^3$D$_{2}$ & 3 & 25 & 1.35e-24 & & & \\
\ion{Fe}{13} & 204.937 & 6.19 & 3s$^2$ 3p$^2$ $^3$P$_{2}$ & 3s$^2$ 3p 3d $^3$D$_{1}$ & 3 & 23 & 1.13e-24 & 8.19 & 0.17 & \\
\ion{Fe}{13} & 246.208 & 6.19 & 3s$^2$ 3p$^2$ $^3$P$_{1}$ & 3s 3p$^3$ $^3$S$_{1}$ & 2 & 14 & 2.35e-24 & 9.36 & 0.18 & \\
\ion{Fe}{13} & 251.953 & 6.19 & 3s$^2$ 3p$^2$ $^3$P$_{2}$ & 3s 3p$^3$ $^3$S$_{1}$ & 3 & 14 & 4.52e-24 & 19.33 & 0.24 & \\
\ion{Fe}{14} & 211.316 & 6.27 & 3s$^2$ 3p $^2$P$_{1/2}$ & 3s$^2$ 3d $^2$D$_{3/2}$ & 1 & 11 & 1.00e-23 & 39.47 & 0.51 & $\heartsuit$\\
\ion{Fe}{14} & 264.787 & 6.27 & 3s$^2$ 3p $^2$P$_{3/2}$ & 3s 3p$^2$ $^2$P$_{3/2}$ & 2 & 10 & 3.72e-24 & 20.72 & 0.08 & \\
\ion{Fe}{14} & 270.519 & 6.27 & 3s$^2$ 3p $^2$P$_{3/2}$ & 3s 3p$^2$ $^2$P$_{1/2}$ & 2 & 9 & 2.56e-24 & 6.96 & 0.06 & $\heartsuit$ \\
\ion{Fe}{14} & 274.203 & 6.27 & 3s$^2$ 3p $^2$P$_{1/2}$ & 3s 3p$^2$ $^2$S$_{1/2}$ & 1 & 8 & 5.54e-24 & 18.31 & 0.08 & $\heartsuit$ \\
\ion{Fe}{15} & 284.160 & 6.33 & 3s$^2$ $^1$S$_{0}$ & 3s 3p $^1$P$_{1}$ & 1 & 5 & 2.88e-23 & 21.20 & 0.14 & ? \\
\ion{Fe}{16} & 262.984 & 6.43 & 3p $^2$P$_{3/2}$ & 3d $^2$D$_{5/2}$ & 3 & 5 & 9.18e-25 & 0.42 & 0.04 & ?
\enddata
\tablenotetext{a}{In this and all subsequent tables wavelengths are in \AA\ and the units
for the intensities and corresponding uncertainties are
erg~cm$^{-2}$~s$^{-1}$~sr$^{-1}$. Lines that can be used for emission measure analysis
are indicated by the ``$\heartsuit$'' symbol. There is no independent check on the
\ion{Fe}{15} 284.160\,\AA\ and \ion{Fe}{16} 262.984\,\AA\ lines so they are marked with a ``?''.}
\label{table:ints1}
\end{deluxetable}
 
\clearpage
 
\begin{deluxetable}{llcr@{\quad---\quad}lrrcr@{\quad$\pm$\quad}rlc}
\tabletypesize{\scriptsize}
\tablewidth{0mm}
\tablecaption{Si, S, and O Intensities Measured in the Quiet Corona above the Limb with EIS}
\tablehead{
\multicolumn{1}{c}{Ion} &
\multicolumn{1}{c}{$\lambda$} &
\multicolumn{1}{c}{$T_{max}$} &
\multicolumn{2}{c}{Transition} &
\multicolumn{1}{c}{L1} &
\multicolumn{1}{c}{L2} &
\multicolumn{1}{c}{$\epsilon_\lambda$} &
\multicolumn{1}{c}{$I_{obs}$} &
\multicolumn{1}{c}{$\sigma_{Iobs}$} &
\multicolumn{1}{c}{}
}
\startdata
\ion{O}{6} & 183.937 & 5.47 & 1s$^2$ 2p $^2$P$_{1/2}$ & 1s$^2$ 3s $^2$S$_{1/2}$ & 2 & 4 & 4.66e-25 & 2.81 & 0.05 & ? \\
\ion{O}{6} & 184.117 & 5.47 & 1s$^2$ 2p $^2$P$_{3/2}$ & 1s$^2$ 3s $^2$S$_{1/2}$ & 3 & 4 & 9.36e-25 & 4.82 & 0.10 & ? \\
\ion{Si}{7} & 275.352 & 5.77 & 2s$^2$ 2p$^4$ $^3$P$_{2}$ & 2s 2p$^5$ $^3$P$_{2}$ & 1 & 6 & 4.55e-24 & 11.54 & 0.07 & $\heartsuit$ \\
\ion{Si}{7} & 275.665 & 5.77 & 2s$^2$ 2p$^4$ $^3$P$_{1}$ & 2s 2p$^5$ $^3$P$_{1}$ & 2 & 7 & 7.04e-25 & 1.81 & 0.05 & $\heartsuit$ \\
\ion{Si}{7} & 278.445 & 5.77 & 2s$^2$ 2p$^4$ $^3$P$_{1}$ & 2s 2p$^5$ $^3$P$_{2}$ & 2 & 6 & 1.45e-24 & 2.95 & 0.19 & $\heartsuit$ \\
\ion{Si}{9} & 258.073 & 6.05 & 2s$^2$ 2p$^2$ $^1$D$_{2}$ & 2s 2p$^3$ $^1$D$_{2}$ & 4 & 13 & 1.31e-25 & 5.08 & 0.08 & $\heartsuit$ \\
\ion{Si}{10} & 253.791 & 6.13 & 2s$^2$ 2p $^2$P$_{1/2}$ & 2s 2p$^2$ $^2$P$_{3/2}$ & 1 & 10 & 4.21e-25 & 12.84 & 0.13 & $\heartsuit$ \\
\ion{Si}{10} & 258.375 & 6.13 & 2s$^2$ 2p $^2$P$_{3/2}$ & 2s 2p$^2$ $^2$P$_{3/2}$ & 2 & 10 & 2.19e-24 & 71.30 & 0.18 & $\heartsuit$ \\
\ion{Si}{10} & 261.058 & 6.13 & 2s$^2$ 2p $^2$P$_{3/2}$ & 2s 2p$^2$ $^2$P$_{1/2}$ & 2 & 9 & 1.30e-24 & 43.34 & 0.13 & $\heartsuit$ \\
\ion{Si}{10} & 271.990 & 6.13 & 2s$^2$ 2p $^2$P$_{1/2}$ & 2s 2p$^2$ $^2$S$_{1/2}$ & 1 & 8 & 9.76e-25 & 37.62 & 0.10 & $\heartsuit$ \\
\ion{Si}{10} & 277.255 & 6.13 & 2s$^2$ 2p $^2$P$_{3/2}$ & 2s 2p$^2$ $^2$S$_{1/2}$ & 2 & 8 & 7.99e-25 & 28.83 & 0.11 & $\heartsuit$ \\
\ion{S}{10} & 264.233 & 6.15 & 2s$^2$ 2p$^3$ $^4$S$_{3/2}$ & 2s 2p$^4$ $^4$P$_{5/2}$ & 1 & 6 & 8.21e-25 & 34.77 & 0.11 & ?
\enddata
\label{table:ints2}
\end{deluxetable}
 
\clearpage
 
\begin{deluxetable}{lcrrc}
\tabletypesize{\footnotesize}
\tablewidth{0mm}
\tablecaption{Isothermal Model Applied to the Si Lines}
\tablehead{
\multicolumn{1}{c}{Ion} &
\multicolumn{1}{c}{Wavelength} &
\multicolumn{1}{c}{$I_{calc}$} &
\multicolumn{1}{c}{$I_{obs}$} &
\multicolumn{1}{c}{$I_{calc}/I_{obs}$}
}
\startdata
\ion{Si}{7} & 275.352 & 11.41 & 11.54 & 0.99 \\
\ion{Si}{7} & 275.665 & 1.83 & 1.81 & 1.01 \\
\ion{Si}{9} & 258.073 & 5.71 & 5.08 & 1.12 \\
\ion{Si}{10} & 253.791 & 14.32 & 12.84 & 1.12 \\
\ion{Si}{10} & 258.375 & 74.43 & 71.30 & 1.04 \\
\ion{Si}{10} & 261.058 & 45.39 & 43.34 & 1.05 \\
\ion{Si}{10} & 271.990 & 34.09 & 37.62 & 0.91 \\
\ion{Si}{10} & 277.255 & 27.89 & 28.83 & 0.97 
\enddata
\label{table:em1}
\end{deluxetable}
 
\clearpage
 
\begin{deluxetable}{lcrrc}
\tabletypesize{\footnotesize}
\tablewidth{0mm}
\tablecaption{MCMC Model Applied to the Fe\,\textsc{ix}--\textsc{xvi} Lines}
\tablehead{
\multicolumn{1}{c}{Ion} &
\multicolumn{1}{c}{Wavelength} &
\multicolumn{1}{c}{$I_{calc}$} &
\multicolumn{1}{c}{$I_{obs}$} &
\multicolumn{1}{c}{$I_{calc}/I_{obs}$}
}
\startdata
\ion{Fe}{9} & 188.497 & 34.13 & 31.28 & 1.09 \\
\ion{Fe}{9} & 189.941 & 18.02 & 15.36 & 1.17 \\
\ion{Fe}{9} & 197.862 & 24.50 & 21.02 & 1.17 \\
\ion{Fe}{10} & 174.532 & 560.17 & 572.57 & 0.98 \\
\ion{Fe}{10} & 177.239 & 304.81 & 308.28 & 0.99 \\
\ion{Fe}{10} & 184.536 & 119.94 & 142.17 & 0.84 \\
\ion{Fe}{11} & 180.401 & 415.93 & 432.24 & 0.96 \\
\ion{Fe}{11} & 182.167 & 53.84 & 58.50 & 0.92 \\
\ion{Fe}{11} & 188.216 & 196.01 & 224.90 & 0.87 \\
\ion{Fe}{11} & 192.813 & 50.16 & 57.73 & 0.87 \\
\ion{Fe}{12} & 186.880 & 33.11 & 35.17 & 0.94 \\
\ion{Fe}{12} & 192.394 & 88.03 & 79.48 & 1.11 \\
\ion{Fe}{12} & 193.509 & 185.60 & 177.53 & 1.05 \\
\ion{Fe}{12} & 195.119 & 277.87 & 274.67 & 1.01 \\
\ion{Fe}{12} & 196.640 & 9.04 & 11.03 & 0.82 \\
\ion{Fe}{13} & 196.525 & 1.71 & 2.71 & 0.63 \\
\ion{Fe}{13} & 197.434 & 5.24 & 7.10 & 0.74 \\
\ion{Fe}{13} & 200.021 & 6.37 & 9.43 & 0.68 \\
\ion{Fe}{13} & 202.044 & 126.16 & 157.66 & 0.80 \\
\ion{Fe}{13} & 203.826 & 27.46 & 25.02 & 1.10 \\
\ion{Fe}{13} & 204.937 & 7.88 & 8.19 & 0.96 \\
\ion{Fe}{14} & 211.316 & 30.71 & 39.47 & 0.78 \\
\ion{Fe}{14} & 270.519 & 7.97 & 6.96 & 1.14 \\
\ion{Fe}{14} & 274.203 & 17.19 & 18.31 & 0.94 \\
\ion{Fe}{15} & 284.160 & 28.72 & 21.20 & 1.35 \\
\ion{Fe}{16} & 262.984 & 0.17 & 0.42 & 0.42
\enddata
\label{table:em2}
\end{deluxetable}
 
\clearpage
 
\end{document}